# Giant strain control of magnetoelectric effect in Ta│Fe│MgO

Dorj Odkhuu

Department of Physics, Incheon National University, Incheon 406-772, Republic of Korea

**ABSTRACT**

**The exploration of electric field controlled magnetism has come under scrutiny for its intriguing magnetoelectric phenomenon as well as technological advances in spintronics. Herein, the tremendous effect of an epitaxial strain on voltage-controlled perpendicular magnetic anisotropy (VPMA) is demonstrated in a transition-metal│ferromagnet│MgO (TM│FM│MgO) heterostructure from first-principles electronic structure computation. By tuning the epitaxial strain in Ta│Fe│MgO as a model system of TM│FM│MgO, we find distinctly different behaviours of VPMA from V- to Λ-shape trends with a substantially large magnetoelectric coefficient, up to an order of 103 fJV$^{-1}$m$^{-1}$. We further reveal that the VPMA modulation under strain is mainly governed by the inherently large spin-orbit coupling of Ta 5$d$–Fe 3$d$ hybridized orbitals at the TM│FM interface, although the Fe 3$d$–O 2$p$ hybridization at the FM│MgO is partly responsible in determining the PMA of Ta│Fe│MgO. These results suggest that the control of epitaxial strain enables the engineering of VPMA, and provides physical insights for the divergent behaviors of VPMA and magnetoelectric coefficients found in TM│FM│MgO experiments.**

# INTRODUCTION

Intensive and extensive attentions have been drawn to transition metal capped ferromagnets on insulating MgO (TM│FM│MgO) as a promising candidate in spin-transfer torque (STT) memory, due to the significant magnetoresistence and perpendicular magnetic anisotropy (PMA), where the spin-polarized current manipulates the spin orientation of a ferromagnet free-layer.[1–3] On the other hand, more recent studies on these magnetic tunnel junctions (MTJs) focus on another alternative approach that utilizes an electric field (E-field) through the voltage to achieve magnetization switching.[4,5] The latter is expected to overcome the large critical current density that is required for STT switching. In particular, a small switching energy per bit, preferably less than 1 fJ, is indispensable, which in STT is in an order of 100 fJ.[6] The large PMA and magnetoelectric (ME) coefficient > 200 fJV$^{-1}$m$^{-1}$ are favored to lower the switching bit energy to 1 fJ in voltage-controlled PMA (VPMA) junctions.[6]

While PMA is discovered in several TM│FM│MgO heterostructures, diverse trends of the E-field dependence of VPMA have been demonstrated ranging from linear to V- to Λ-shape behaviors. For example, Ta│CoFeB│MgO and Au│CoFe│MgO exhibit the linear VPMA with magnetoelectric coefficients of –33 and –38 fJV$^{-1}$m$^{-1}$, respectively.[7,8] On the other hand, nonlinear Λ- and V-shape VPMA have also been reported when Fe-rich layers are used as a FM constitute in V│Fe│MgO[9] and MgO│FeB│MgO,[10] which have significantly large magnetoelectrics of –1150 and 108 fJV$^{-1}$m$^{-1}$, respectively. There have been numerous subsequent studies since then to clarify the origin of such a wide range of VPMA and ME coefficients. However, the underlying mechanism remains unclear. It is indeed challenging to straightforwardly answer within a theoretical model because of the limited understanding on a detailed atomic structure, composition, and possible diffusion of boron into MgO.[3] Apparently, even the physics origin of PMA in TM│FM│MgO multilayers is still not fully understood. The pioneering studies, for instance, have stated that the observed PMA of Ta│CoFeB│MgO is contributed predominantly by the interface layer between the FeCoB and MgO.[2,4] In recent experimental[11,12] and theoretical studies,[13–15] the role of TM capping layers on PMA has also been addressed and cannot be ignored, due to the inherently large spin-orbit coupling (SOC) of 4d and 5d orbitals. The latter studies claimed that the presence of the TM│FM interface is an essential ingredient in determining the PMA,[11–14] broadening the common understanding that the Fe 3$d$–O 2$p$ hybridization is the main origin.[2,4,16,17]

One of the main causes for different behaviours of VPMA could be an epitaxial strain at the interfaces, either TM│FM and/or FM│MgO, as lattice mismatch occurs quite often in thin films that are epitaxially grown. The validity of this argument has been addressed in a very recent experiment, where an unexpectedly large VPMA modulation of up to 7000 fJV$^{-1}$m$^{-1}$ was reported in strained Ta│CoFeB│MgO films.[18] Such a significant modification of VPMA under strain should be correlated to the strain- and E-field-induced shifts of the relative occupations of d orbitals,[19] where MA energy (MAE) is determined by the SOC interaction between occupied and unoccupied bands as[20]

$$MAE = \xi^2 \sum_{o,u} \frac{|\langle o|l_z|u\rangle|^2 - |\langle o|l_x|u\rangle|^2}{\varepsilon_u - \varepsilon_o}, \qquad (1)$$

where $o$ ($u$) and $\varepsilon_u$ ($\varepsilon_o$) represent the eigenstates and eigenvalues of occupied (unoccupied) states, respectively, and $\xi$ is the strength of SOC interaction. For unstrained bulk Ta and Fe lattices, where c/a = 1, the crystal field under $O_h$ point group splits five-fold degenerate $d$ orbital states into doublets ($e_g$) and triplets ($t_{2g}$). When the lattice changes from cubic-symmetry c/a = 1 to low-symmetry c/a < 1 under an epitaxial strain (in the presence of MgO), as schematically shown in Fig. 1 for the case of Fe│MgO interface, the crystal field splitting ($e_g$ into $d_{z^2}$ and $d_{x^2-y^2}$, and $t_{2g}$ into $d_{xy}$, $d_{xz}$, and $d_{yz}$) may offer more freedom to provide more energy differences in the denominator of Eq. (1).[19] It is therefore important to see how the different energy levels of $d$ orbitals in different crystal symmetry evolve upon an E-field in order to identify the most relevant physics of VPMA.

Herein, we present results of first-principles calculations to show a decisive role of a strain effect on E-field dependence of VPMA in TM│FM│MgO multilayers. While an asymmetric V-shape E-field dependence of VPMA is demonstrated in Ta│Fe│MgO for the unstrained in-plane lattice of MgO, the compressive strain gives rise to the (lambda)-shape behavior with a substantially large ME coefficient, up to an order of 1000 fJV$^{-1}$m$^{-1}$. These are the results of the strain- and field-induced changes of the strong spin-orbit coupled Ta 5$d$-orbitals hybridized with the Fe 3$d$-orbital states at the TM│FM interface, and may provide physical insights for the divergent behaviors of VPMA and ME coefficients that are observed in experiments.[7–10,18]

## RESULTS

It is known that in most TM│FM│MgO multilayers the film thickness of TM and FM layers, especially FM, plays an important role in determining the MAE. In experiments, for example, the typical thickness of FM layers that exhibit PMA in Ta/CoFeB/MgO is within the range of 0.5–1.2 nm.[2,21] In accord with a realistic situation, we first explore the effect of the Fe thickness ($n$), ranging $n$ = 3–6 atomic layers or about 0.5–1 nm, on magnetism and MAE in Ta$_3$│Fe$_n$│(MgO)$_5$, where the numbers in the subscript denote the number of atomic layers. Considering the lack of generality and reality, the results corresponding to the Ta and Fe thickness at $n$ = 1–2 are excluded. As shown in Fig. S1 in Supplementary Information (SI), the saturation behavior of magnetic moments at the interfaces is evident as the number of Fe layers increases just beyond the $n$ = 3 and PMA is preferred for all $n$ although it shows an oscillatory behaviour as a function of thickness, analogs to the Fe│MgO films in previous theoretical[22,23] and experimental studies.[24,25] Here and hereafter, we thus refer the results to those in Ta$_3$│Fe$_3$│(MgO)$_5$, whose geometry is illustrated in Fig. 2(a). However, we would like to note that the direct comparison with measurements from the broad ranges of sample stoichiometry and film-thickness requires some cautious, and the further calculations of the effects of strain and E-field on MAE with more geometric patterns, including thicker films of FM and TM layers, should be carried out to provide more insights. From the total energy calculations, the preferred adsorption sites of Fe atoms were atop of O and a hollow site at the Fe|MgO and Ta|Fe interface, respectively. The magnetic moments and interlayer distances at the interface between Fe and O, denoted as $d_{Fe-O}$, and Fe and Ta ($d_{Fe-Ta}$), for a different strain $\eta$ are shown in Table 1. Hereafter, we refer to zero strain ($\eta = 0$) as the experimental lattice constant (2.978 Å) of MgO, while $\eta = -4\%$ corresponds to the lattice constant (2.86 Å) in bulk Fe. We recall here that the in-plane lattice near the Fe|MgO interface in practical is within this range, $-4\% \leq \eta \leq 0$.[26] The in-plane lattice of Ta capping is then enforced to match accordingly for each strain, assuming a few-nanometers-thick TM and FM films. Both the $d_{Fe-O}$ and $d_{Fe-Ta}$ decrease as $\eta$ increases from $-4\%$ to zero. The magnetic moments of the interface Fe atoms, labeled as Fe(1) at the Fe|MgO and Fe(2) at the Ta|Fe interface (Fig. 2(a)), exhibit a trend similar to the interlayer distances, which is predominantly due to the charge transfer and hybridization effects at the interfaces.[13] For the unstrained lattice of MgO ($\eta = 0$), the Fe(1) (Fe(2)) atoms have an enhanced (reduced) moment of 2.67 (1.50) $\mu_B$ in the presence of MgO (Ta) layers with respect to the center Fe or bulk feature of pristine Fe. The interface Ta atom has a noticeable induced moment (–0.32 $\mu_B$), antiparallel to the Fe moments.

These changes associated with the hybridization effect are well manifested in the minority spin density of states (DOS) that is shown in Fig. 2(b). The presences of MgO and Ta layers severely affect the $d$ states of Fe in bulk (See Fig. 4 in Ref. 27); While a large peak appears above the Fermi level in the Fe(1)-DOS, the DOS of the Fe(2) atom are rather dispersive and shifted downward across the Fermi level, which can be inferred as the reflection of the enhanced (reduced) spin exchange-splitting of the Fe(1) (Fe(2)) site through the Fe $d$–O $p$ (Fe $d$–Ta $d$) hybridization. Furthermore, these electronic states near the Fermi level are altered significantly as one shrinks in-plane lattice from $\eta = 0$ (blue) to $\eta = -2\%$ (red) to $\eta = -4\%$ (black line in Fig. 2(b)). In particular, the coincidence of simultaneous shifts of Fe(2)- and Ta-DOS towards the high energy level is prominent. As a consequence, the interfacial Fe (Ta) moments increase (decreases) as strain increases, and tend to achieve their free-standing surface magnetism (nearly 3 $\mu_B$ for Fe and zero for Ta atom) at $\eta$ = −4%. The changes of the fully occupied majority spin states under different strains are not significant, so they are not seriously considered in the present study.

In Fig. 2(c), the strain dependence of the zero-field MAE of Ta│Fe│MgO is shown for $\eta$ = −4% to zero. While PMA is preserved for all $\eta$ considered (which is not the case for Ta│CoFe│MgO[28]), the $\eta$ = −2%, which is the compressive (tensile) strain relative to the MgO (Fe) lattice, is found to have a greatly enhanced MAE of 2.5 erg/cm$^2$. This value is notably larger than that (about 0.5 erg/cm$^2$) found in strained Ta│CoFe│MgO with $\eta$ = −2%,[26] which along with the robust PMA under different strains is worth noting from a practical viewpoint to prevent a stable magnetization axis from thermal fluctuations.[3] The contribution of two interfaces to MAE is also analyzed from the individual systems of Fe│MgO without Ta (left in Fig. 2(a)) and Ta│Fe without MgO layers (middle in Fig. 2(a)). To see the relative contribution of interfaces, the half of MAE value of the three atomic layers of free-standing Fe(001) is subtracted from the total MAE for each interface. The strain-induced enhancement of MAE, $\Delta$MAE($\Delta\eta_1/\eta_2$) = MAE($\eta$ = −2%) − MAE($\eta$ = −4%/0), is presented in Fig. 2(d) for Fe│MgO (orange), Ta│Fe (violet), and Ta│Fe│MgO (dark gray). Notably, both the $\Delta$MAE($\Delta\eta_1$) and $\Delta$MAE($\Delta\eta_2$) of the Ta│Fe interface exhibit a trend similar or comparable in the magnitude of MAE to those for Ta│Fe│MgO, but Fe│MgO does not. This indicates that the dominant contribution to the variation of MAE value under strain comes mainly from the Ta│Fe interface, leading to the lambda ($\Lambda$)-shape trend shown in Fig. 2(c). The previous *ab initio* calculations have already shown that the large changes of MAE in magnetic metals alloyed or capped with 4*d* and 5*d* elements are the results of the larger SOC of 4*d* and 5*d* orbitals than the 3*d* elements.[13,29] On the other hand, the intrinsic nature of the PMA of Ta│Fe│MgO can be attributed to the interface effects at the Fe│MgO: Fe *d*–O *p* hybridization[2,4,16,17] and perfect epitaxy of Fe films on the MgO surface.[23] Furthermore, the strain affects not only the magnitude of MAE but also the charge transfer considerably at the Ta|Fe interface rather than at the Fe|MgO, which is the result of the interfacial charge screening effects due to the internal E-field across the interface.[30] Indeed, this can be modulated via strain, and such strain-induced modification of the charge transfer evolves in changes of the SOC bands near the Fermi level, which will be correlated with MAE in the forthcoming discussion.

To get more insights, we further decompose MAE on $k$-space, MAE($k$), over a two-dimensional (2D) Brillouin zone (BZ) in Fig. 3(a) for $\eta$ = –4%, –2%, and zero (from left to right), where the blue/dark (yellow/light) area represents negative (positive) MAE($k$). The dominant contribution that yields the larger PMA at $\eta$ = –2%, with respect to those for $\eta$ = –4% and zero strains, is mainly along the BZ line between 2/5(ΓX) and 2/5(XM), as indicated by the dashed-line in Fig. 3(a). For better visualization, $\Delta$MAE($k,\Delta\eta_1$) and $\Delta$MAE($k,\Delta\eta_2$), as defined before, along 2/5(ΓX)–2/5(XM) are also plotted in Fig. 3(b). The corresponding band structures in the spin-down state of the Ta atom at the Ta|Fe interface are shown in Figs. 3(c), (d), and (e) for $\eta$ = –4%, –2%, and zero strain, respectively. For the spin-channel contributions to the MAE, we follow a recipe by the previous full-potential calculations on TM|Fe (TM=Ru, Rh, Pd, Os, Ir, and Pt) systems to which the spin down-down (↓↓) channel contributes dominantly over the other spin channels, spin up-down ↑↓ and up-up ↑↑.[13] We decompose the Eq. (1) into nonvanishing matrix elements with the SOC eigenvalue states that are predominant near the Fermi level in the ↓↓-component, where the SOC constant is omitted, as

$$MAE = \xi^2 \sum_{o,u} \frac{\left|\langle o|l_z|u\rangle\right|^2 - \left|\langle o|l_x|u\rangle\right|^2}{\varepsilon_u - \varepsilon_o}, \quad (2)$$

where, the positive and negative contributions to MAE are determined by $l_z$ and $l_x$ operators, respectively.[20]

The band analyses will be concentrated on particular $k$-points, at and around 2/5(ΓX), in which the dominant peak contributions of ΔMAE($k$,Δ$η_1$/$η_2$) (Fig. 3(b)) and the eigenvalue states near the Fermi level (Fig. 3(d)) are prominent. As η increases from –4% to zero, the following two features become notable: (1) the filled $d_{xy}$ and $d_{xz}$ (empty $d_{yz}$) bands shifts upward (downward) above (towards) the Fermi level, (2) the partially filled $d_{x^2-y^2}$ band becomes fully occupied and appears just below the Fermi level at $η$ = –2% and moves further away from the Fermi level when $η$ = 0. For $η$ = –4% at 2/5(ΓX), the hybridized $d_{xy}$ and $d_{xz}$ bands coexist below and above the Fermi level, of which the occupied $d_{xz}$ band also couples with the $d_{yz}$ at around 0.5 eV. These couplings give rise to the competition of SOC states between the second positive and third negative terms in Eq. (2). When $η$ = –2%, however, substantial rearrangements of occupied and unoccupied bands near the Fermi level, as (1) and (2), provide additional positive contributions to the first and second terms in Eq. (2), through $\langle xz|l_z|yz \rangle$ and $\langle x^2-y^2|l_z|xy \rangle$. The former and latter positive contributions are reduced significantly at $η$ = 0 because of the absence of filled $d_{xz}$ band and the enlarged energy difference in the denominator of the second term in Eq. (2), respectively. To support the aforementioned atomic and electronic origin of PMA changes under strain, we also analyze the energy- and $k$-resolved band distribution of orbital characters of the interface Fe(1) and Fe(2) atoms (See the SI). In contrast to the Ta and Fe(2) at the TM│FM interface, there are no appreciable coupling states that can give a contribution to positive ΔMAE(Δ$η_1$/$η_2$) at 2/5(ΓX) for the Fe(1) atom at the Fe│MgO interface. Only the SOC pairs between the states that yields the negative contribution by $\langle x^2-y^2|l_z|xy \rangle$ exist around 2/5(ΓX) at $η$ = –2%, as shown in Fig. S2 in the SI.

We now explore the effects of E-field and strain on VPMA. The external E-field is oriented normal to the in-plane lattice, where the upward direction pointing toward the Fe│MgO interface represents positive the E-field, as depicted in Fig. 1. The dipole correction was taken into account to eliminate an artificial field across the slab imposed by the periodic boundary condition. The variation of the interfacial Fe(1), Fe(2), and Ta site magnetism as a function of E-field is presented in Fig. S3 in the SI for $\eta$ = –4%, –2%, and zero strain. Here, we refer the effective E-field to that in MgO, which would be more practical than the local fields at the interface, defined as E-field(MgO) = E-field(vac)/$\varepsilon$, where E-field(vac) is an external E-field in vacuum and $\varepsilon$ is the dielectric constant of MgO. The calculated (experimental) $\varepsilon$ are about 20, 12.5, and 10 (9.8) for $\eta$ = –4%, –2%, and zero, respectively.[28,31] From the slope of VPMA with respect to the E-field in MgO, one can determine the ME coefficient by VPMA/E-field(MgO). In addition to the modification of E-fields with strain, the external E-field is also expected to decay at each interface due to the different work functions between the two materials in the stacked layers.

Figures 4(a)–4(c) show the E-field dependence of MAE of Ta│Fe│MgO for $\eta$ = –4%, –2%, and zero strain, respectively. Being in the opposite trend to the E-field dependence of Ta magnetism except for $\eta$ = –4% (Fig. S3 in the SI), distinctly different behaviours of VPMA are demonstrated for different strains. The E-field dependence of VPMA at $\eta$ = –4% exhibits an asymmetric V-shape trend with a minimum at –0.25 V/nm similar to the magnetism, where ME coefficients of about –1210 and 600 fJV$^{-1}$m$^{-1}$ are calculated for E-field < 0 and E-field > 0, respectively. On the other hand, at $\eta$ = –2% the negative field reduces MAE monotonically while MAE increases with the positive field up to 0.4 V/nm, and then sharply decreases with further increases in field (Fig. 4(b)). This leads to the $\Lambda$-shaped VPMA with 169 (–746) fJV$^{-1}$m$^{-1}$ for the negative (positive) field, which is opposite that for $\eta$ = –4%. As $\eta$ further goes to zero, V-shaped VPMA reappears with a local minimum at the negative field, as in $\eta$ = –4%. The smaller dielectric constant (10) of MgO for the zero strain causes the reduction of the ME coefficient, –270 (449) fJV$^{-1}$m$^{-1}$ when E-field < 0 (E-field > 0). Such different behaviours of VPMA and critical fields corresponding to the maximum and minimum MAE values can be ascribed to the interplay between the internal and external E-fields

through Rashba splitting of the band structure. A recent study reported that the Rashba SOC interaction, which has contributions from both the internal and external E-fields, can give an important contribution to the MAE with a nonlinear quadratic form.[30] This indicates that the strain- (internal E-field) and external E-field-induced shifts of the *d* orbitals in energy levels, as will be clarified later, lead to the aforementioned results. Note that these ME constants are two to three times of the corresponding values found in Ta│CoFe│MgO,[28] and, in general, are larger than the ~200 fJV$^{-1}$m$^{-1}$ required to achieve a switching bit energy below 1fJ.[6]

The calculated results suggest that the VPMA and ME coefficient can be sensitively controlled by engineering the in-plane lattice. We also emphasize here that the strain effect might be the main cause of experimental deviations in the VPMA and ME coefficient, as mentioned previously.[7–10,18] For instance, V│Fe and MgO│FeB interfaces, which differ in the interface structure as well as the capping layers, exhibit the opposite V- and Λ-shaped VPMA but also ME coefficients in an order of discrepancy.[9,10] In particular, with $\eta$ = –4% we reproduce the experimental magnetoelectric coefficient of –1150 ± 50 fJV$^{-1}$m$^{-1}$ found in V│Fe│MgO, which was attributed to the E-field-induced charge screening at the interfaces.[9] In this junction, a thinner MgO layer (1.2 nm) is sandwiched by 0.7- and 5-nm-thick Fe layers,[9] in which the in-plane lattice of MgO is most likely enforced to adjust to that of Fe during epitaxial growth. More recently, an importance of epitaxial strain on enhancement of VPMA was revealed in ferromagnetic resonance measurements on Ta│CoFeB│MgO films.[18]

To observe the contribution of the interface layers to the variation of VPMA under the E-field, we calculate the MAE as a function of the E-field for the individual Ta│Fe and Fe│MgO interfaces at $\eta$ = –4% (Fig. S4 in the SI). As seen in Fig. S4(a), the Ta│Fe interface exhibits the V-shape E-field dependence of MAE similar to that (Fig. 4(a)) of Ta│Fe│MgO, even though the magnitude of MAE is more and less zero or negative under the E-field. In contrast, the VPMA is rather insensitive in the presence of the E-field for the Fe│MgO interface. It was reported in experiments that the presence of MgO is necessary for the PMA in Ta│CoFeB│MgO from the fact that Ta│CoFeB│Ta without MgO layers exhibit in-plane magnetization. The weak or negative MAE at the Ta│Fe interface can be explained by the 3$d$–5$d$ hybridization and the band-filling effects, which were discussed in detail in Refs. 13 and 26. As mentioned before, the PMA at the Fe│MgO interface is due to the Fe $d$–O $p$ hybridization[2,4,16,17] and perfect epitaxy of Fe films on the MgO surface.[23] Thus, although the Fe│MgO interface is responsible for determining the PMA,[2,4,16,17,23] the variation of MAE under the E-field in Ta│Fe│MgO is predominantly contributed by the Ta│Fe interface (Fig. S4(b) in the SI).[15]

In order to elucidate the different trends of VPMA and ME coefficients under strain, the distribution of the E-field-induced change of MAE($k$), $\Delta$MAE($k$) = MAE($\pm$E,$k$) – MAE(0,$k$), over full BZ is shown in Figs. 4(d)–4(f) for $\eta$ = –4%, –2%, and zero, respectively. The +E and –E denote the direction of the E-field (i.e., E-field > 0 and E-field < 0). The magnitudes of $\pm$E are $\pm$1, $\pm$1.25, and $\pm$1.5 V/nm for $\eta$ = –4%, –2%, and zero, respectively. For each $\eta$, we will focus on the particular $\Delta$MAE($k$) region that provides a dominant contrition to VPMA, as shown in the bottom panels in Figs. 4(d)–4(f): around 1/4(XM) along 1/4($\Gamma$X)–1/4(XM) for $\eta$ = –4%; around 1/3($\Gamma$X) along 1/3($\Gamma$X)–1/3(XM) for $\eta$ = –2%; and around 3/5($\Gamma$X) along the 3/5($\Gamma$X)–3/5(XM) line for zero strain. We also plot the corresponding zero-field band structure of orbital characters of the interface Ta in Figs. 5(a)–(c), and the shifts under E-field < 0 (blue solid line) and E-field < 0 (red solid line) in Figs. 5(d)–(f) for $\eta$ = –4%, –2%, and zero, respectively. At $\eta$ = –4%, the downward (upward) shift of the filled $d_{yz}$ (unfilled $d_{x^2-y^2}$) under E-field < 0 (E-field > 0) around 1/4(XM) reduces negative contribution ($\langle x^2 - y^2 | l_z | yz \rangle$) in MAE, resulting in the positive peaks in Fig. 4(d). The other sharp positive peak above 1/4($\Gamma$X) when E-field < 0 is ascribed to the shift of the empty $d_{x^2-y^2}$ band towards the Fermi level, which couples with the occupied $d_{xy}$ just below the Fermi level. This positive $\Delta$MAE(–

E,*k*) through $\langle x^2-y^2|l_z|xy\rangle$, which is not present in E-field > 0, leads to the larger VPMA slope when the E-field < 0 (−1210 fJV$^{-1}$m$^{-1}$) in Fig. 4(a).

At *η* = −2%, the zero-field negative contribution by $\langle xy|l_x|xz\rangle$ around 1/3(ΓX) (Fig. 5(b)) is significantly increased in the presence of the E-fields since the downward (upward) shift of empty $d_{xz}$ (filled $d_{xy}$) band enhances the negative third term of Eq. (2), as shown in Fig. 5(e). From this, the negative peaks of ΔMAE(*k*) at 1/3(ΓX) are well addressed in Fig. 4(e), thereby ΔMAE(*k*) < 0 for E-field < 0 and > 0. On the other hand, at *η* = 0 the zero-field positive MAE should be partially contributed by the SOC pair states between the $d_{xz}$ and $d_{yz}$ bands across the Fermi level at 3/5(ΓX) (Fig. 5(c)). Under the fields, as seen in Fig. 5(e), the ΔMAE(±E,*k*) around 3/5(ΓX) should increase due to the reduced energy difference between the $d_{xz}$ and $d_{yz}$ states in the denominator of the first term in Eq. (2).

## DISCUSSION

Through the first-principles electronic structure calculations, we identify a synergistic effect of epitaxial strain on PMA and its E-field dependence of VPMA in Ta│Fe│MgO multilayers. By tuning an in-plane strain, we find a substantially large magnetoelectric coefficient, in an order up to 1000 fJV$^{-1}$m$^{-1}$. We further provide the physical insights on the origin of different behaviors of V- and Λ-shaped VPMA, where the large SOC of the magnetically-induced 5*d* Ta atoms through the strong Ta 5*d*–Fe 3*d* hybridization at the TM│FM interface plays an important role although the Fe 3*d*–O 2*p* hybridization at the Fe│MgO interface is partly responsible in determining the PMA of Ta│Fe│MgO as addressed previously. In the context of spintronics applications, the present study suggests that strain engineering would be a highly desirable route to achieve the large voltage-induced modulation of perpendicular magnetic anisotropy required in magnetoelectric memory devices.

# METHODS

Density functional theory (DFT) calculations were performed using the projector augmented wave (PAW) pseudopotential method,[32] as implemented in the Vienna *ab initio* simulation package (VASP).[33] Exchange and correlation interactions between electrons were described with the generalized gradient approximation (GGA) formulated by Perdew, Burke, and Ernzerhof (PBE).[34] Energy cutoff 500 eV and Monkhorst-Pack k-mesh of 15 × 15 × 1 were imposed for the ionic relaxation, where forces acting on atoms were less than $10^{-2}$ eV/Å. The SOC term is included by using the second-variation method employing the scalar-relativistic eigenfunctions of the valence states.[35] MAE is calculated from the total energy difference when the magnetization directions are on the *xy*-plane ($E^{\parallel}$) and along the *z*-axis ($E^{\perp}$), MAE = $(1/a)^2 \cdot (E^{\parallel} - E^{\perp})$, where *a* is the in-plane lattice constant, so that positive MAE stands for the preferable direction of magnetization normal to the film plane (i.e., PMA). A dense k-point of 31 × 31 × 1 was used for MAE calculations, which was sufficient to get reliable results.

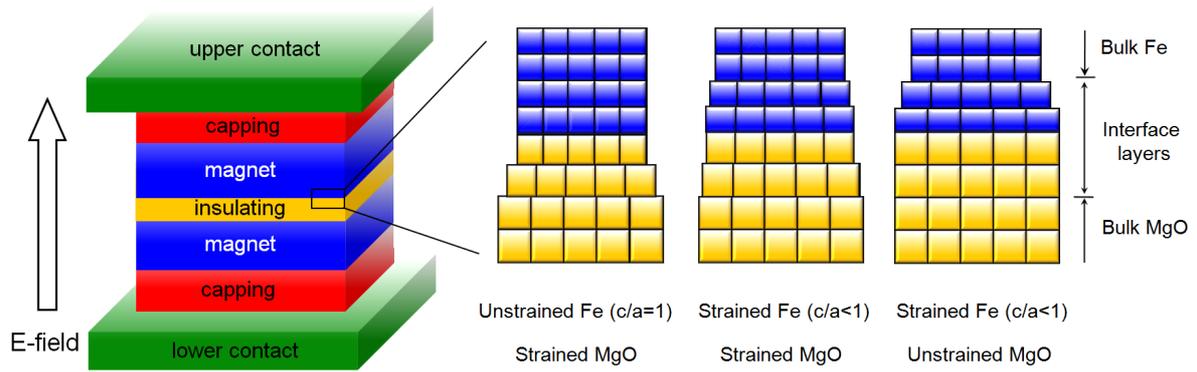

**Figure 1.** (a) Schematics for the in-plane lattice matching at the interface between Fe (blue) and MgO (yellow) of Ta│Fe│MgO: Compressive strain on MgO lattice (left), expansive strain on Fe lattice (right), or either at the same time under an epitaxial condition (center). The upward arrow in the right indicates the direction of the positive electric field.

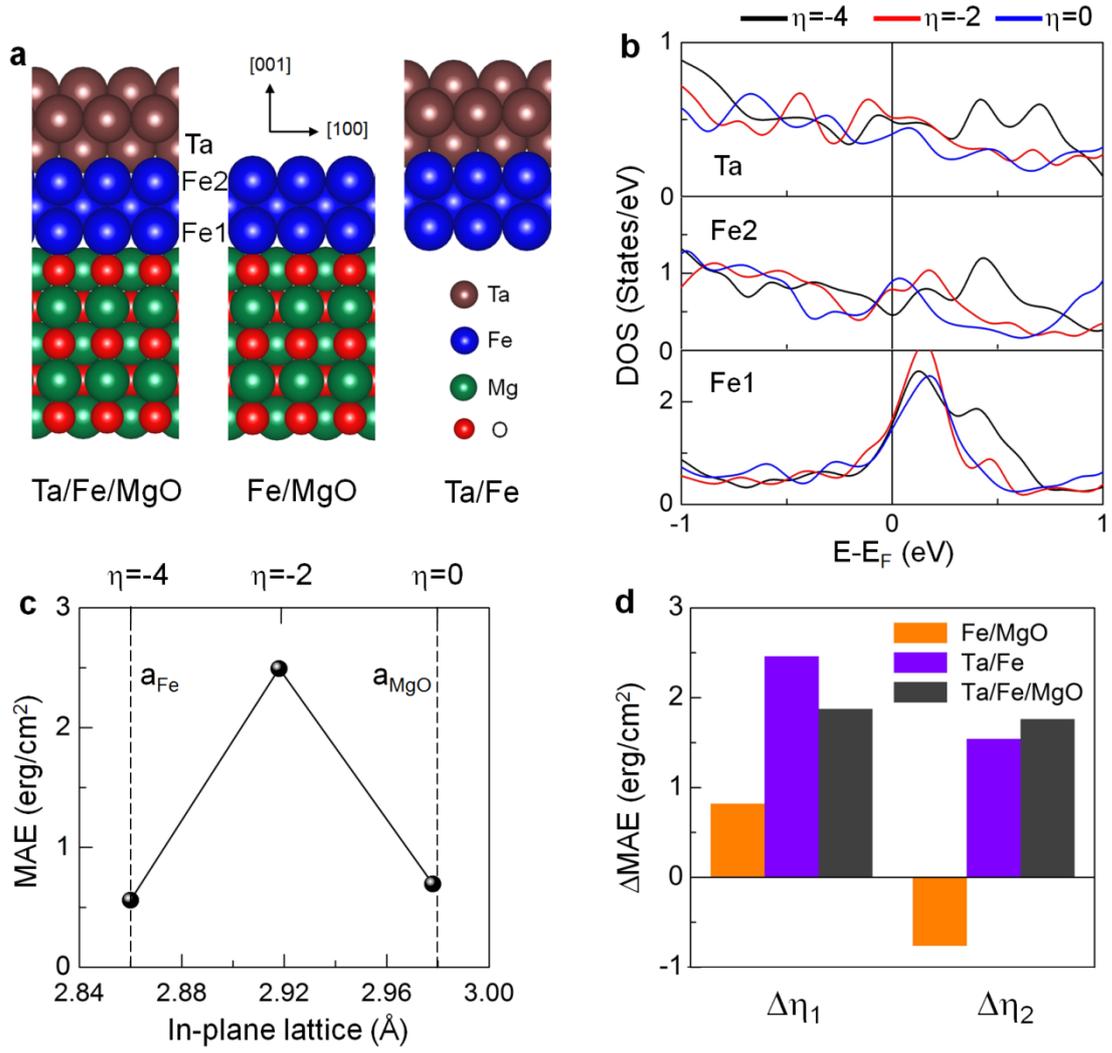

**Figure 2.** (a) Side view of atomic structures for Ta│Fe│MgO (left), Fe│MgO without Ta (center), Ta│Fe without MgO layers (right). The larger brown, blue, green, and smaller red balls are Ta, Fe, Mg, and O atoms, respectively. The Fe atoms at the Fe│MgO and Ta│Fe interfaces are labeled by Fe(1) and Fe(2), respectively. (b) Density of states in the minority-spin state of the Ta (topmost), Fe(2) (center), and Fe(1) (bottommost panel) atoms for different strains of $\eta$ = –4% (black), –2% (red), and zero (blue). The Fermi level is set to zero energy. (c) MAE of Ta│Fe│MgO for $\eta$ = –4%, –2%, and zero strain. The vertical dashed lines indicate the experimental lattice constants of bulk Fe and MgO, which correspond to $\eta$ = –4% and zero strain, respectively. (d) Strain-induced changes in MAE, $\Delta$MAE($\Delta\eta_1/\eta_2$) = MAE($\eta$ = –2%)–MAE($\eta$ = –4%/0), for the individual Fe│MgO (orange) and Ta│Fe (violet area)

interfaces. Results of ∆MAE(∆$\eta_1$/$\eta_2$) for Ta│Fe│MgO are also shown in dark-gray for comparison.

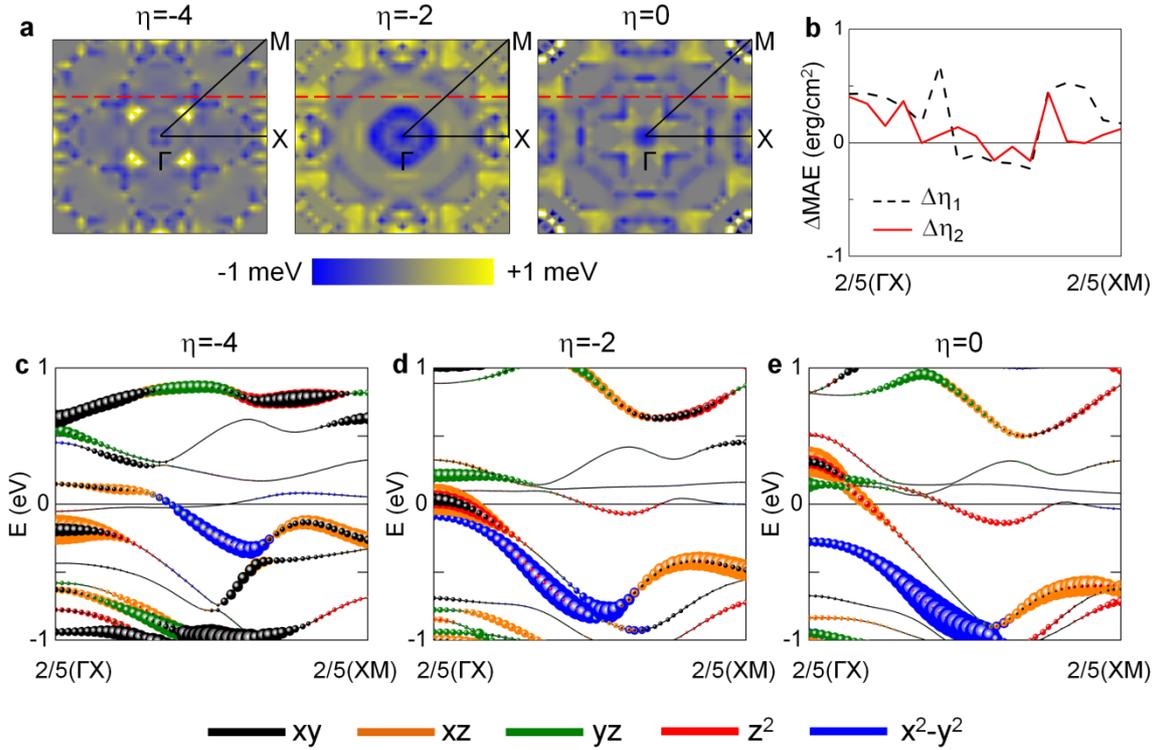

**Figure 3.** (a) Distribution of MAE over *k*-space, MAE(*k*), for $\eta$ = –4%, –2%, and zero strain (left to right). Blue/dark and yellow/light areas represent negative and positive MAE(*k*), respectively. The horizontal dashed lines indicate the *k*-point line that can represent the strain-induced changes of total MAE. (c) MAE(*k*) difference along 2/5(ΓX)–2/5(XM) between at $\eta$ = –4% and $\eta$ = –2% (dashed), and between at $\eta$ = 0 and $\eta$ = –2% (solid). Minority-spin band structures along the 2/5(ΓX)–2/5(XM) line for (c) $\eta$ = –4%, (d) $\eta$ = –2%, and (e) zero strain. The $d_{xy}$, $d_{xz}$, $d_{yz}$, $d_{z^2}$, and $d_{x^2-y^2}$ orbital characters of Ta atom at the Ta│Fe interface are denoted in black, orange, green, red, and blue, respectively. The symbol size represents the weight of the *d* orbitals. The Fermi level is set to zero energy.

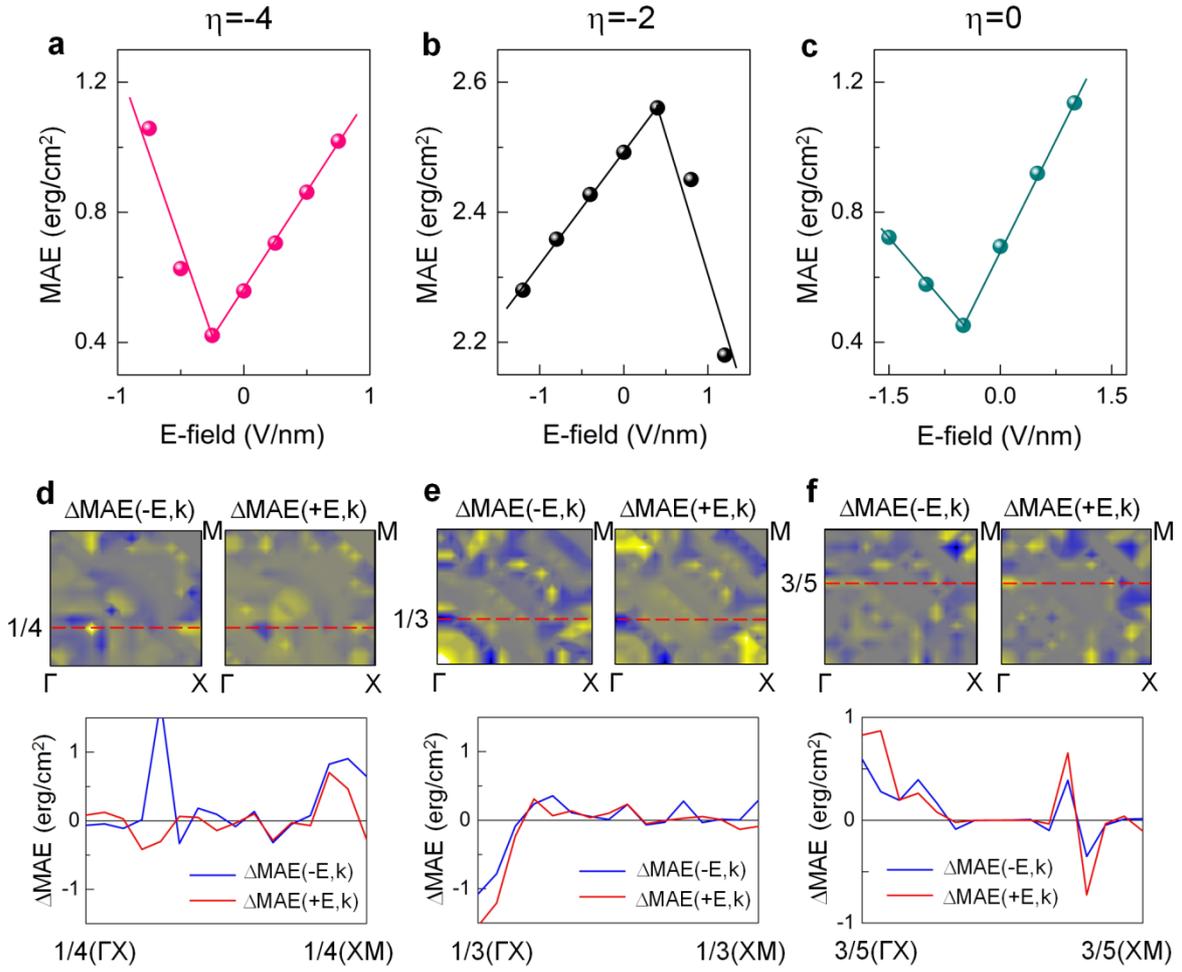

**Figure 4.** Electric-field dependence of MAE for (a) $\eta$ = –4%, (b) $\eta$ = –2%, and (c) zero strain. The data points with a circle symbol are fitted by a linear in all panels. The value of the electric field in the horizontal axis is that in MgO. Electric-field-induced MAE($k$), $\Delta$MAE($k$) = MAE($\pm E,k$) – MAE(0,$k$), for negative (left) and positive field (right) for (d) $\eta$ = –4%, (e) $\eta$ = –2%, and (f) zero strain. Symbols (– and +) denote the direction of the electric field. For each strain, $\Delta$MAE($\pm E,k$) along the particular k-point line, which has the largest contribution to MAE, as indicated by the horizontal dashed line, is detailed in the bottom panels of (d)–(f) for negative (blue) and positive electric-field (red).

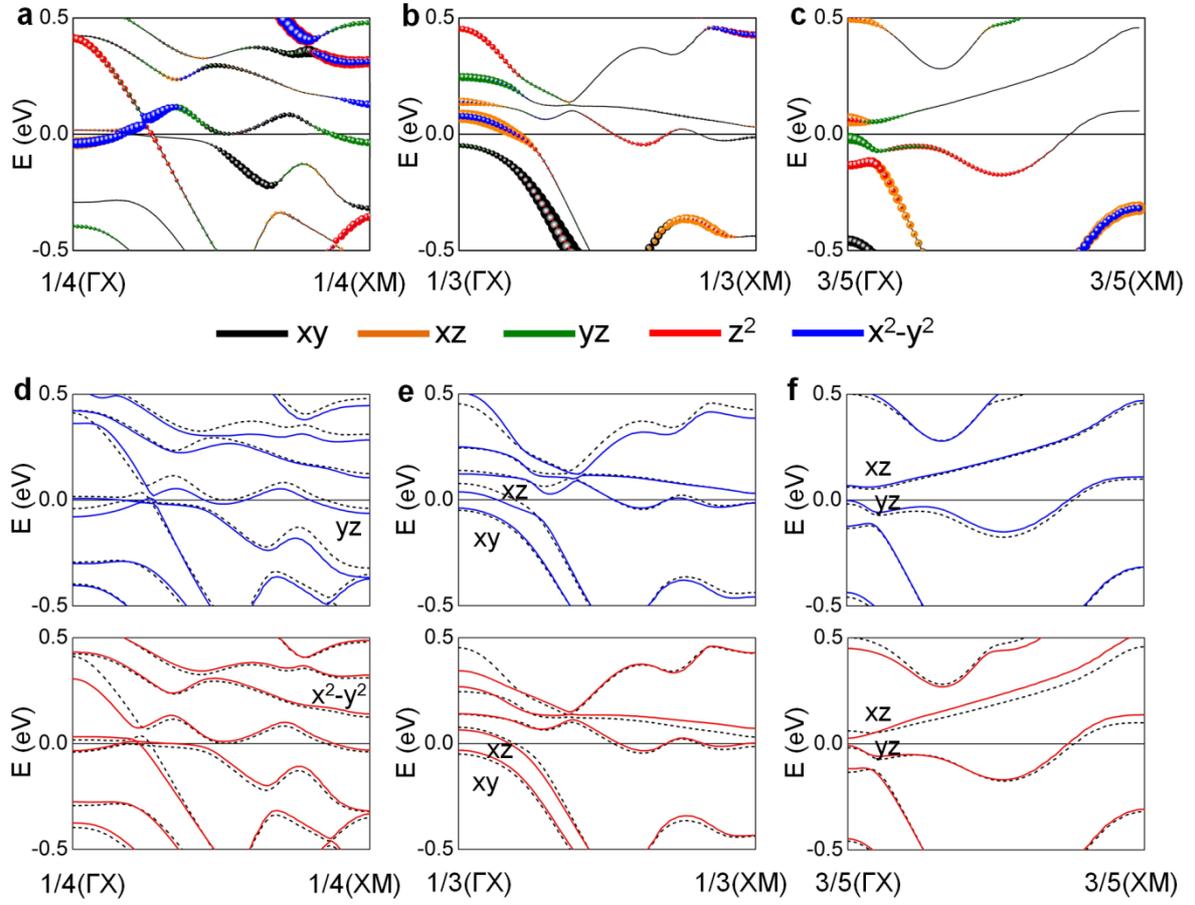

**Figure 5.** Minority-spin band structures along the particular *k*-point line shown in Figs. 3(d)–3(f) for (a) $\eta = -4\%$, (b) $\eta = -2\%$, and (c) zero strain. $d_{xy}$, $d_{xz}$, $d_{yz}$, $d_{z^2}$, and $d_{x^2-y^2}$ orbital characters of Ta atom at the Ta│Fe interface are denoted in black, orange, green, red, and blue, respectively. The symbol size represents the weight of the *d* orbitals. Electric-field-induced shifts of band structures for negative (blue) and positive electric-field (red) are shown for (d) $\eta = -4\%$, (e) $\eta = -2\%$, and (f) zero strain. For comparison, zero field band structures are also shown in the dotted black lines. The Fermi level is set to zero energy.

**Table 1.** Optimized interlayer distances (Å), and magnetic moments ($\mu_B$) of Fe(1) at the Fe|MgO and Fe(2) and Ta atoms at the Ta|Fe interface of Ta|Fe|MgO for different strains in the zero electric field. Strain at the experimental lattice constant (a = 2.987 Å) of MgO is taken as reference. The largest compressive strain with $\eta$ = –4% considered in this study corresponds to the experimental lattice constant (a = 2.86 Å) of bulk Fe.

| $\eta$ | a | $d_{Fe-O}$ | $d_{Fe-Ta}$ | $M_{Fe(1)}$ | $M_{Fe(2)}$ | $M_{Ta}$ |
|---|---|---|---|---|---|---|
| –4% | 2.860 | 2.21 | 1.79 | 2.862 | 2.146 | –0.068 |
| –2% | 2.918 | 2.13 | 1.51 | 2.676 | 1.517 | –0.266 |
| 0 | 2.978 | 2.08 | 1.45 | 2.670 | 1.502 | –0.316 |

# Supplementary Information

# Giant strain control of magnetoelectric effect in Ta/Fe/MgO

## Dorj Odkhuu

Department of Physics, Incheon National University, Incheon 406-772, Republic of Korea

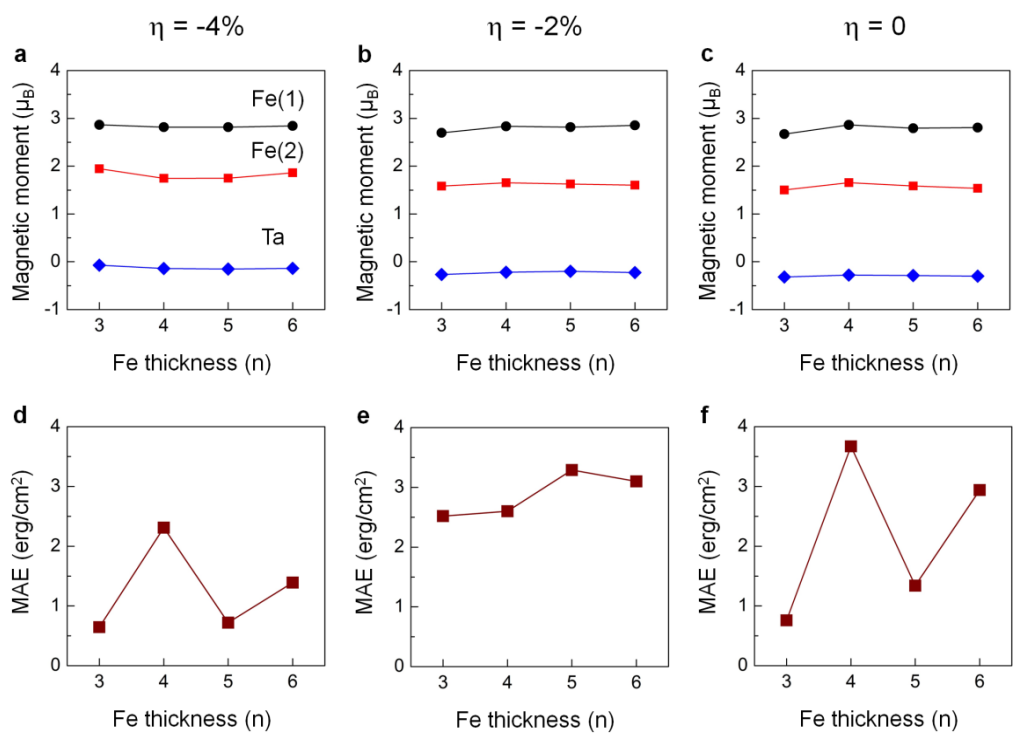

**Figure S1.** Fe-thickness-dependent magnetic moments of Fe(1) (circle) at the Fe/MgO, and Fe(2) (square) and Ta (diamond) atoms at the Ta/Fe interface of Ta/Fe/MgO for (a) $\eta$ = –4%, (b) $\eta$ = –2%, and (c) zero strain. Fe-thickness-dependent MAE of Ta/Fe/MgO for (d) $\eta$ = –4%, (e) $\eta$ = –2%, and (f) zero strain.

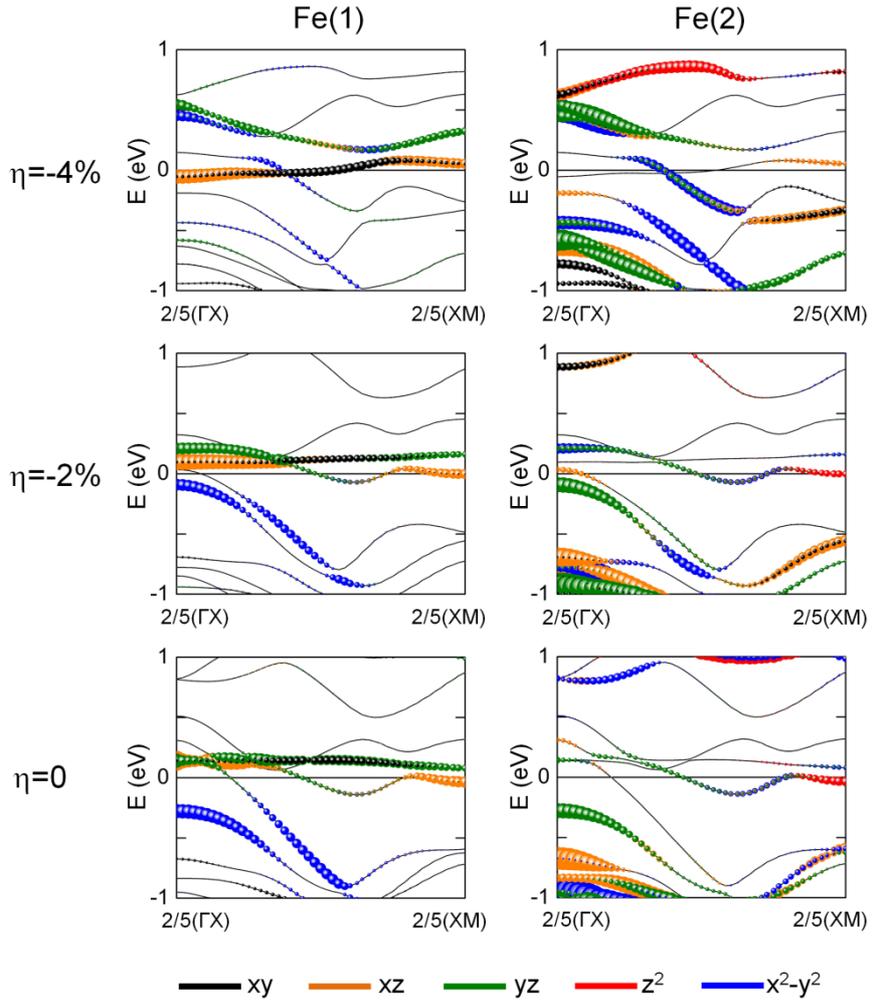

**Figure S2.** Minority-spin band structures along the 2/5(GX)–2/5(XM) line of Fe(1) at the Fe/MgO interface and Fe(2) at the Ta/Fe interface of Ta/Fe/MgO for $\eta = -4\%$, $\eta = -2\%$, and zero strain. The orbital states $d_{xy}$, $d_{xz}$, $d_{yz}$, $d_{z^2}$, and $d_{x^2-y^2}$ are denoted in black, orange, green, red, and blue, respectively. The symbol size represents the weight of the $d$ orbitals. The Fermi level is set to zero energy.

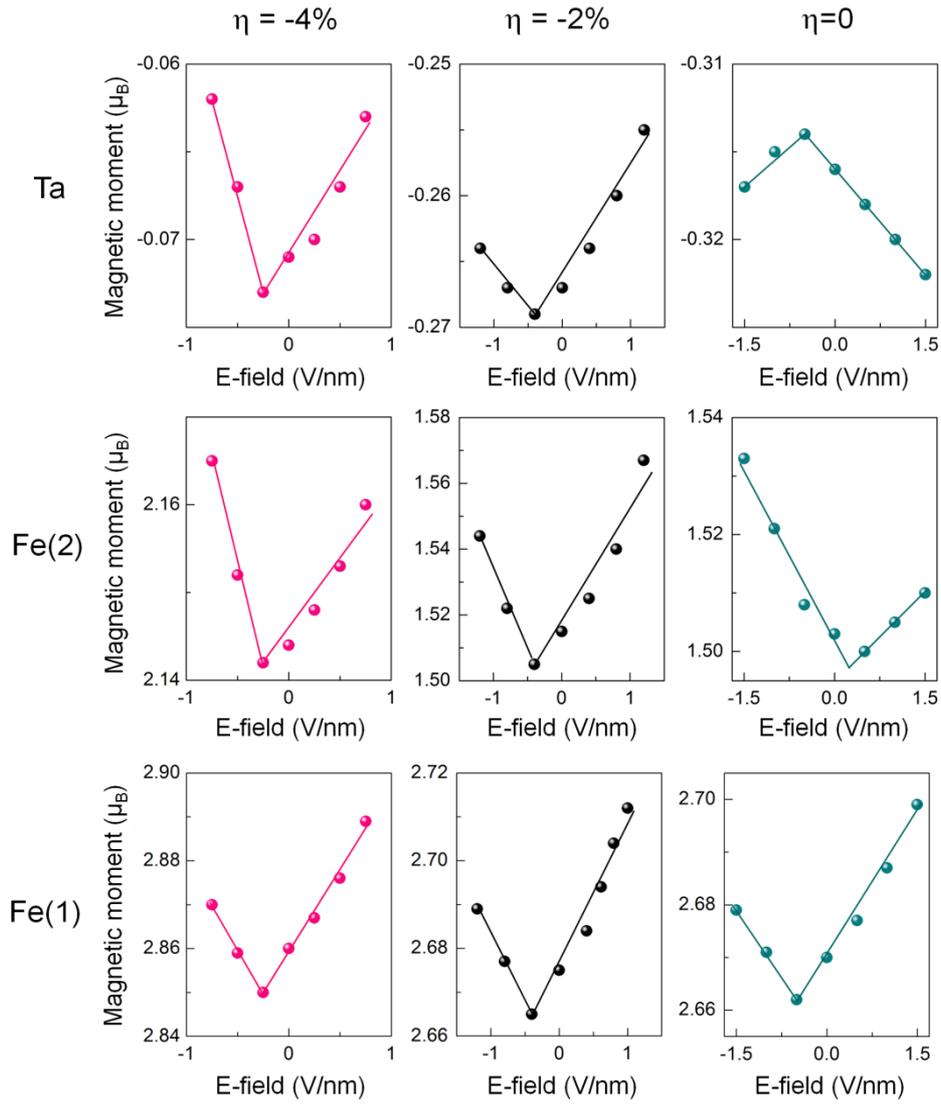

**Figure S3.** Electric-field dependence of Ta (topmost), Fe(2) (middle), and Fe(1) (bottommost) moments of Ta/Fe/MgO for $\eta$ = –4% (left), $\eta$ = –2% (center), and zero strain (right). The data points with circle symbol are fitted by a linear in all panels. The value of electric-field in the horizontal axis is that in MgO.

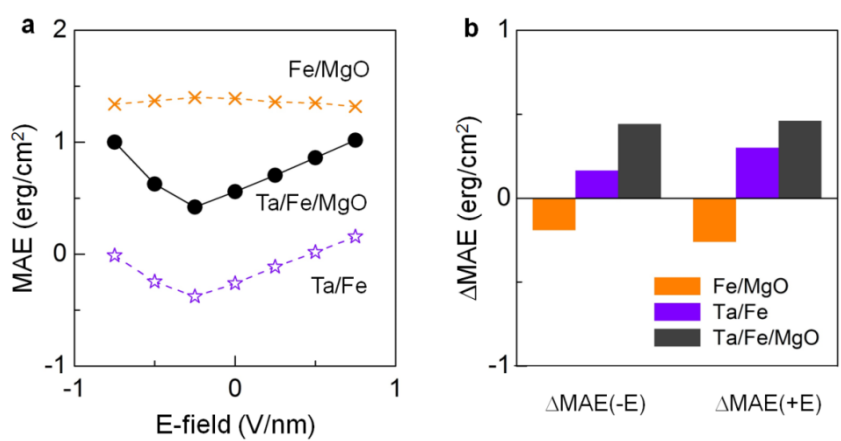

**Figure S4.** (a) Electric-field dependence of MAE for the individual Fe/MgO (orange cross) and Ta/Fe (violet star) interfaces. Results for Ta/Fe/MgO are also shown in black circles for comparison. (b) Electric-field-induced MAE, ΔMAE(±E) = MAE(±E) – MAE(0), for the individual Fe/MgO (orange) and Ta/Fe (violet area) interfaces. Results for Ta/Fe/MgO are also shown in dark-gray area for comparison.